# Independent components in spectroscopic analysis of complex mixtures

**Yulia B. Monakhova [*a] , Sergey A. Astakhov [a], Alexander Kraskov [b], Svetlana P. Mushtakova [a]**

[a] *Department of Chemistry, Saratov State University, Astrakhanskaya 83, Saratov, Russia 410012*

[b] *UCL Institute of Neurology, London, WC1N 3BG, UK*

## ABSTRACT

We applied two methods of "blind" spectral decomposition (MILCA and SNICA) to quantitative and qualitative analysis of UV absorption spectra of several non-trivial mixture types. Both methods use the concept of statistical independence and aim at the reconstruction of minimally dependent components from a linear mixture. We examined mixtures of major ecotoxicants (aromatic and polyaromatic hydrocarbons), amino acids and complex mixtures of vitamins in a veterinary drug. Both MICLA and SNICA were able to recover concentrations and individual spectra with minimal errors comparable with instrumental noise. In most cases their performance was similar to or better than that of other chemometric methods such as MCR-ALS, SIMPLISMA, RADICAL, JADE and FastICA. These results suggest that the ICA methods used in this study are suitable for real life applications.

## INTRODUCTION

Quantitative spectroscopic methods have offered widely used analytical tools while themselves being a subject of active research and development. These methods open a window into exploring complex mixtures by combining some unique features, including a unified approach (many of them treat various spectral signals – e.g. IR, UV, visible, scattering, EPR, time-resolved – in a single manner), non-destructive measurements (especially important for *in vivo* analysis of natural samples) and an opportunity to separate measurements and data processing.

The most challenging problems in mixture decomposition are the estimation of the number of most significant mixture components, identification of the chemical nature of the components and estimation of their concentrations. When only mixed spectral signals are available, analysts speak of a "black mixture" (drawing a line to the "black box"[1]). However, *a priori* information about the mixture composition and/or further assumptions act to "whiten" the problem, formalize it and make numerical predictions feasible. We will now review what different approaches offer along these lines.

---
[*] Corresponding author.
*E-mail adress:* yul-monakhova@mail.ru (Yu. Monakhova), tel.: +78452516953, fax:+78452516950.

Having experimental spectra of pure mixture constituents at hand reduces quantitative mixture analysis to linear algebra. Evidently, there are some considerable difficulties on that route. Indeed, routine production of standard samples has proved to be cumbersome and expensive. Besides, now it is almost impossible to make a set of standard samples to meet the wide and growing range of suspected compounds that typically fall under analysis. For instance, as many as about 700 major contaminants are standardized in the course of water and air control [2]. Obviously, even in this case forming a reliable database of high-quality standards would hardly be realistic. This has triggered research into alternatives.

Methods for molecular modeling and theory of spectra suggested a key to standardless mixture analysis [2,3]. A breakthrough was made by substituting natural samples by their mathematical counterparts, i.e. calculated spectra of molecular models for the presumed mixture constituents. These approaches look quite promising in environmental analysis where standard samples make almost no practical sense. As of today chemists recognize more than 10 million individual chemical substances. Taking persistent molecular transformations into account, the actual number is unknown and is growing.

Furthermore, it turned out that the use of time-resolved spectra [3-5] opens up an opportunity to make mixture analysis completely free of standards (based on experimental data only). This has been shown possible due to the information presented by the dynamics of spectral lines [6-8].

The above two variants that originate from computational spectroscopy have several of essential features in common, and namely they: (i) make qualitative assumptions about the mixture constituents and (ii) predict the mixture composition (quantitative analysis) through modeling the analytical signal (computing intensity distributions in wavelength and/or time for molecular models).

In parallel, a representative arsenal of more than 20 alternative and competing methods for mixture analysis has been developed over past 30 years in chemometrics (see the reviews, e.g., in [1,9-18]). These methods are focused to seek a solution to the "black mixture" problem by resorting to an abstract mixture model as superposition of unknown components with no assumptions about their molecular structure or type of spectra. The chemometrics methods pioneered in spectral analysis by Lawton and Sylvestre [19] are targeted at making quantitative predictions about the concentrations and pure component spectra (self-modeling solution).

Mixture systems often contain molecules with similar or identical structural groups. Plus, in certain cases analysis is complicated by extensive spectral details (e.g., vibration structure in UV spectra of polyaromatic hydrocarbons, PAHs). Chemical analysis of such mixtures is a nontrivial task, which stimulates searches for efficient techniques. One of the approaches developed in chemometrics is based on identification of "pure variables" for all mixture components [20]. A pure variable is a frequency (wavelength) at which the contribution from one of the components dominates. The pure variables, thus, approximately mark the regions where at least one of the spectral components is independent from all others.

The principal difficulty here is the manifold of self-modeling solutions (including some physically meaningless ones such as those with negative concentrations or non-positive spectra). To guarantee obtaining unique solutions, several constraints and empirical criteria have been proposed (non-negativity, assumed spectral shapes, closure, choice of active spectral bands, etc.). Also, the resulting decomposed pure spectra are, in a sense, abstract and require identification (either automated through spectral databases search or by experts). In contrast with molecular modeling and standard sample techniques, here quantification (i.e. blind reconstructions of abstract mixture components and their concentrations) goes prior to qualitative analysis (identification of chemical structures).

In a wider context, decomposition of arbitrary mixed signals into pure components has received considerable attention and remains the subject of extensive research in signal processing with numerous applications in telecommunications, geophysics, image processing, bioinformatics, medicine. This set of approaches, generally termed "blind source separation" (BSS), has its most developed branch known as "independent component analysis" (ICA) [21, 22]. Basic ICA solves the decomposition problem assuming linearity of the mixture and statistical independence of individual stationary signals. In some interpretations, ICA can be considered as an extension of principal component analysis (PCA), a building block of many chemometrics methods. ICA methods differ in numerical measures of statistical independence and approximations. Recently, the focus in ICA research has shifted to improving performance and accuracy. One of these high fidelity universal ICA techniques – MILCA (Mutual Information Least Dependent Component Analysis [23]) – is employed in the present study. MILCA is based on the search for least dependent (in contrast to independent) mixture components gauged by precise numerical estimates of mutual information [24] as a measure of signal dependence.

There have been noticeable evidences that universal BSS/ICA methods merge actively into the field of applications in spectroscopy and chemometrics. As the literature surveys indicate, the number of practical analytical problems solved by these methods has increased at least tenfold over the last decade. In analytical spectroscopy, BSS/ICA can be classified as standardless and self-modeling techniques. Equally important is also the reverse – specifics of spectral experiment suggest optimal algorithms for independent component analysis [25]. One of the algorithms used in this paper, SNICA (Stochastic Non-Negative Independent Component Analysis [26]), was motivated by spectroscopy applications. It naturally combines efficient decomposition through minimization of mutual information between components and non-negativity constraint. The latter is characteristic to many types of experimental spectral data. Building upon statistically representative ensembles of synthetic and experimental mixtures, numerical experiments have shown that MILCA and SNICA outperform specialized chemometrics and other ICA algorithms on typical blind source separation problems, including spectroscopic problems [23,25,26]. This article examines the proposed methods in analytical practice. It reports on a series of case studies which evaluate the performance of MILCA and SNICA methods on real problems judged by the quality criteria and standards adopted in chemistry. The mixtures under analysis differ in mixture design, spectral bandwidths, the extent of band overlaps and wavelength counts. These factors together with user experience allow evaluating the method from a practical analytical perspective.

**COMPUTATIONAL METHODS**

In a common with chemometrics formulation, the linear ICA problem can be presented as:

$$\mathbf{X} = \mathbf{A}\,\mathbf{S},\qquad\qquad(1)$$

where $\mathbf{X}$ is a M×N matrix of M measured mixture spectra, $\mathbf{S}$ is a K×N matrix of K unknown spectra of pure components (here N denotes the number of counts over wavelength), $\mathbf{A}$ is a M×K mixture matrix (unknown concentrations). The task is to reconstruct $\mathbf{S}$ and $\mathbf{A}$ (prediction) given the observed $\mathbf{X}$, assuming that the original pure components are mutually as independent as possible (hypothesis constraining the manifold of solutions).

Then the idea of ICA application to the mixture decomposition is based on the following considerations. The pure component spectra show only weak dependences (although they do not have to be assumed strictly independent in the statistical sense [24,25]). These dependences will be stronger if the components are similar. Mixing make the observed signals $\mathbf{X}$ more dependent than the pure sources $\mathbf{S}$. ICA then seeks a transformation that "compensate" for the dependences caused by mixing, i.e. a decomposition matrix $\mathbf{W}$ (the resulting estimate for $\mathbf{A}^{-1}$) such that it minimizes interdependencies in $\mathbf{Y} = \mathbf{W}\,\mathbf{X}$ (the resulting estimates for $\mathbf{S}$). ICA relies on mutual information $I(\mathbf{Y})$ as a quantitative measure of statistical dependence. Signals (rows of $\mathbf{Y}$) are considered statistically independent and have zero mutual information if their joint distribution factorizes into individual distributions.

The basic MILCA method [23] uses precise numerical estimates for mutual information based on a nearest neighbors algorithm [24] (parameterized by the number of nearest neighbors $K_{nn}$) and makes no assumptions about the individual distributions of source (pure) signals. SNICA is a method dedicated to analysis of non-negative signals and performs best on signals with intensity distributions peaked at zero (the case typical in spectroscopy). Like ALS (alternating least squares), SNICA has non-negativity constraints. The non-negativity constraint in tandem with the "minimal dependence" assumption made it possible to exclude the sometimes detrimental PCA preprocessing altogether [25,26] and achieve better performance in the case of dependent pure components. The method is build upon Monte-Carlo minimization and simulated annealing to avoid spurious solutions. The essential SNICA parameters [26] are the number of nearest neighbors $K_{nn}$, Monte Carlo initial step size $h_0$, "temperature" T and stopping criterion M. The efficiency of ICA methods on spectral data can be considerably improved by performing decomposition in derivative space (derivatives of spectral curves with respect to wavelength using finite differences, smoothing Savitzky-Golay filters, splines [23, 25,26]). ICA solutions $\mathbf{Y}$ and $\mathbf{W}$ are scale and permutation invariant that is why only relative concentrations and spectral curves in relative units make physical sense while analyzing ICA outputs.

In a real analytical experiment the true concentrations and components are unknown which calls for empirical criteria or indirect indications of decomposition performance. The following results may be considered indicative of incomplete decomposition: a combination of positive and negative resulting concentrations for some components, alternating sign spectral curves, slow convergence of Monte-Carlo optimization and $\mathbf{W}$ close to unit matrix (in case of SNICA), numerical noise in the output, significant variance of mutual information of components recovered by different methods or with respect to small variations of method parameters for a fixed mixture. For some analytically difficult mixtures a combination of decomposition methods or Monte-Carlo-like optimization of method parameters may be advantageous.

One should realize however that the ICA methods are not entirely standardless. The decomposition techniques do not require training data sets or specialized methods for reconstruct matrices S and A. In essence, ICA methods provide a "blind" decomposition to resolve spectra and the relative concentration profiles of all coexisting substances. But one needs standard samples and/or library spectra to identify resolved compounds and compute their concentration profiles in physical units. There are some alternatives to solve this problem depending on available resources and analyst's preferences.

It is commonly accepted that mixtures under analysis fall into three groups: white, grey and black [1]. The white mixtures are the easiest one because information about all chemical species and coexisting interferents is available. Quantitative analysis of that sort of systems is now well-grounded. Black mixtures, as mentioned in the previous section, are those for which no information is available. In such cases once the resolved spectra obtained qualitative analysis can be performed based on library spectra (for example, NIST database [27]) or spectra of expected substances recorded in laboratory. The later is preferable because reference spectral signal should be taken in one experimental setting. This technique is widely used in this paper.

Further, in the quantification step, one needs samples of known concentration for the resolved chemical components. Here all methods developed for white systems are suitable (one can try MLR or PLS methods). Also analyst can determine concentrations by self-modeling decomposition of spectral data of objects with standard additions of substances under analysis. The basis of the method is the comparison responses of a mixture sample of unknown concentrations with the responses obtained after adding standards of known concentrations. This method is very useful when matrix effect should be evaluated. Likewise, one can quantify only one compound in the mixture by alternative technique and then reproduce other abundances. In addition, in cases where total concentration is known quantitative analysis can rely only on relative ICA concentrations. Possible ways to verify results include the method of standard additives, comparison with other chemometrics techniques or reference method, if it exists. Thus, the strategy on quantitative step for black systems should ideally transforms black system into a white or gray.

In this article concentrations of components in absolute (physical) units in model mixtures were calculated using the known total molar concentration of the systems and relative ICA concentrations. For the real object («Nitamin» drug), the actual concentrations were determined by decomposition of medicine spectra with known standard additives of all vitamins.

MILCA, SNICA, and the mutual information algorithm are implemented as standalone executables for Windows/Linux and also have MATLAB interfaces. The download packages along with the GPL open sources are available for free at the website [28]. To compute matrix inverse we used the Moore-Penrose method and LAPACK algorithms [29] (MATLAB implementations). We also made some experimental data used in this paper available at [28].

**EXPERIMENTAL**

The UV-VIS spectra were recorded at 0.1, 0.2, 0.5, 1 nm resolution on Cary-100, HP8452A and SHIMADZU-1800 spectrometers with the cells having path-lengths of 0.2 and 1 cm. It is necessary that the number of mixture spectra must be equal or greater than the number of compounds in the mixture.

For our study we selected chemical systems analysis of which is often a rather complicated task. First, we studied multicomponent mixtures of organic substances that are very important in environmental analysis, specifically, aromatic compounds (benzene, toluene, xylene) and polyaromatic hydrocarbons. Further, we investigated two classes of biologically active substances – vitamins and amino acids. Complex mixtures of vitamins represent a class of organic compounds which have to be mixed and than checked with great precision. Separation and analysis of amino acids and their derivatives are often used in determination of amino acid composition of complex proteins, peptide sequencing and in diagnostics of metabolic disorders.

Currently the determination of above mentioned compounds is carried out using chromatographic techniques, voltamperometric and potentiometric methods. However, time constraints and other limitations make these techniques unsuitable for routine analysis. One problem that hinders a wider application of fluorescence and absorption spectroscopy in environmental monitoring is the lack of selectivity of spectroscopic measurements caused by strongly overlapping bands of these compounds [30].

Thus, the problem of simultaneous determination of specified substances in complex objects is a really actual issue. Chemometrics approaches presented in this study can offer a solution to this problem and open a window into simple, reliable and fast exploration of complex systems.

All materials and solvents used were of analytical-reagent grade. Solutions of PAHs ($10^{-3}$ M) were prepared by weighing and transferring appropriate amount of reagents to 25 ml volumetric flasks, dissolved and completed to volume with hexane. Working solutions of benzene, toluene, o-xylene were made by dissolving appropriate volume of reagents in hexane. To prepare samples for registration, further dilutions from each stock solution were made using hexane.

Solutions of vitamins (B6, B9 and B12) for analysis of model systems were prepared by transferring 0.5 g of each compound to 100 ml volumetric flasks, adding 1 ml of triethanolamine and dissolving in distilled water. Then we adjusted pH to 7.5 with 10% solution of hydrochloric acid. Solutions for recordings were made by dissolving working solution with distilled water. Solutions of C, PP, B6 vitamins ($10^{-2}$ mol/l) were prepared by weighing, dissolved and completed to volume with hydrochloric acid (0.1 mol/l).

Working solution of a veterinary drug «Nitamin» was made by dissolving 1 ml of the original medicine in 100 ml volumetric flask with distilled water. Vitamin C sample was made by dissolving 0.5 g of substance in 100 ml volumetric flask with distilled water. Solutions of vitamins E and A were prepared by dissolving accurate amount of 0.25 and 0.294 g respectively in Solutol HS15 and then dissolved in distilled water in volumetric flasks of 50 and 100 ml respectively. Recorded spectra were adjusted to pH 6-80-6.90 with 10% solution of hydrochloric acid and 10% sodium hydroxide.

Working solutions of amino acids were prepared by weighing appropriate amount of compound and then dissolving in bidistilled water.

Detailed description of all data sets (resolution, number of points) is provided in Table I.

For all calculations we use Matlab v. 7.0 (The Math Works, Natick, MA, USA) and PLS-Toolbox v. 5.2 (Eigenvector research, Wenatchee, WA, USA). Computational time in all cases under consideration was below 10 minutes per system (mixture) including preparation of the samples.

To characterize similarity between experimental and calculated matrixes we apply Amari index:

$$P_{err} = \frac{1}{2N} \sum_{i,j=1}^{N} \left( \frac{|p_{ij}|}{\max_k |p_{ik}|} + \frac{|p_{ij}|}{\max_k |p_{kj}|} \right) - 1, \qquad (2)$$

where $p_{ij} = (\hat{A}^{-1}A)_{ij}$.

The Amari index vanishes when the recovered concentrations differ from the true ones only in scaling and permutation of components, and it increases as the quality of decomposition becomes poor. Thus, small values of the Amari index are desirable. (In practice, we find that good decomposition quality roughly corresponds to Amari indices $P < 0.05$, whereas $P > 0.2$ generally characterizes unacceptably poor performance).

We compared well-known and freely available MCR-ALS [31, 32], SIMPLISMA [21, 33] methods as well as JADE [34], RADICAL [35] and FastICA [36] algorithms with MILCA and SNICA. In the present study we use resolved MILCA spectra as initial estimates for MCR-ALS. During the ALS optimization, we applied non-negativity constraints to model the shapes of both the spectra and concentration profiles.

To assess the similarities between the normalized resolved and the original experimental (pure) spectra, we use frequently applied correlation coefficient (R). Normalized spectra were obtained by dividing every absorbance on the maximum value per spectrum.

Before recording spectra for decomposition the influence of different factors (noise, step of spectral scan and scan speed) on decomposition performance has been investigated and the optimal conditions for spectroscopic registration have been identified. Specifically, 0.5-2 nm spectral resolution and Medium speed scan were found to give sufficient accuracy of analysis together with short registration time for our systems.

All experiments were repeated three times, the tables report mean values together with standard deviations. The details of decomposition runs (derivative space where used, the values of $K_{nn}$, scan step) are attached to the tables with quantitative results.

**RESULTS**

*Aromatic compounds (benzene - toluene - xylene)*

We selected two- and three-component mixtures with various concentrations of benzene, toluene and o-xylene for analysis. Experimental spectra of ternary benzene-toluene-o-xylene mixtures are shown on figure 1a. MILCA was used to perform decomposition. The results of qualitative (Fig. 1b) and quantitative (Table 1) analyses indicate a very good fit with the "ground truth" data. The relative quantification error in concentrations was 12 % for ternary systems, the locations of absorption bands were determined with the 0.5 nm accuracy, which is comparable with instrumental errors (the worst spectrum estimate was achieved for benzene (R=0.96)). Original experimentally measured sources and obtained IC are shown in Fig 6a.

*Polyaromatic hydrocarborns (PAHs)*

To examine the applicability of ICA in qualitative and quantitative analysis of PAHs we decomposed various multicomponent mixtures with different composition (up to 6 substances). The following compounds were chosen in our analysis: anthracene (1), pyrene (2), phenantrene (3), benz[a]antracene (4), benz[a]phenantrene (5), fluorantene (6), 1-aminopyrene (7), 1-bromopyrene (8), 1-pyrenecarboxyaldehyde (9). Mixtures were diversified in the number of components and composition.

We study anthracene-pyrene–phenantrene system as a mixture of three typical PAHs representatives. The recovered and experimental spectra are almost identical with the relative error in locations of absorption peaks less than 0.5 nm and values of correlation coefficients are not less than 0.96. Table 1 shows that the relative error in concentration for ternary mixtures is below 5%.

We also tested four- and six- component mixtures. As an example, the experimental spectra of four-component anthracene-phenantrene-benz[a]antracene-benz[a]phenantrene system along with the results of qualitative analysis are shown on Fig. 2. Table 1 demonstrates the results of quantitative analysis of the above mentioned and some other systems of PAHs with varying composition. The quantification uncertainties were below 10 %, which is indication of suitability and robustness of MILCA algorithm on this class of mixtures and molecular structures.

Next, an important issue of ICA decomposition is the determination of the concentration range of all substances which one can reliably quantify with given accuracy. We probed a concentration range (ranging from 1:15 to 15:1 in ratio) and evaluated decomposition accuracy on binary pyrene–phenantrene system. We found that reliable qualitative analysis can be performed even for mixtures with extreme concentration ratios. Quantitative analysis manifests some explainable accuracy degradation at low and high concentration ratios where also instrumental errors may be significant. Our data indicate that simultaneous quantitative analysis is possible for mixtures in concentration interval $1.3 - 20 \cdot 10^6$ M with relative error below 15%.

*Complex mixtures of vitamins*

The analysis of multicomponent vitamin mixtures without chemical separation of constituents is a rather difficult task. Derivative spectrometry has been used for the mixture analysis of three vitamins with overlapping spectra (B1-B6-B12) by zero-crossing measurements [37]. However, sometimes the derivative techniques cannot cope with the level of interference especially when the spectra are strongly overlapped or in case of complex mixtures or where mixture composition is unknown.

We examine mixtures of vitamin pairs and triplets. As an example, experimental spectra and the results of qualitative analysis of the three-component system B6-B9-B12 are shown on Fig. 3. It can be observed that calculated and experimental spectra correlate well. Quantitative analysis is reported in Table 1 with the relative error in concentrations below 8%.

Approbation of ICA decomposition algorithms on real objects is of great practical interest, especially where mixture composition is not known exactly. Here we apply MILCA to analysis of vitamins in a complex multivitamin veterinary drug «Nitamin». The medicine contains a mixture of three vitamins (A, E and C) in a complex matrix (12 compounds in total). This fact makes quantitative chromatographic analysis very difficult.

We registered the spectra of solutions prepared by mixing the appropriate quantities of the medicine with standard addition of all vitamins (Fig. 4a). The pH level was kept constant in order to prevent spectral band shifts. Spectra of individual vitamins were extracted (Fig.4 b) and the concentrations of compounds in original medicine were obtained within 10% relative error (Table 2).

Typically, analysis of vitamins in complex mixtures by ICA methods took about 5 minutes. Given the accuracy achieved, this makes ICA decomposition an attractive express analytical alternative.

*Amino acids*

In this study we analyzed binary and ternary mixtures of amino acids which have strong absorption in UV region (tyrosine and tryptophan) and creatinine (the final product of protein metabolism) in the wide concentration range (Table 1). All mentioned compounds can be found in one biological object. Fig. 5 and Table 1 contain both quantitative and qualitative results of analysis of selected compounds. Maximum relative error was found to be 11 %. Original experimentally measured sources and obtained IC are shown in Fig 6b.

**DISCUSSION**

In the previous section we have demonstrated that the pure component spectra of the constituents in complex mixtures could effectively be recovered using the new ICA algorithms. Correlation coefficients close to one were obtained between estimated and reference spectra even though sources had highly overlapping spectral bands (for examples, in case of aromatic compounds) (Fig.1-5). Additionally, as Fig.6 demonstrates, the estimated IC's spectra are similar to the "ground truth" data and can be used in identification of mixture component. It was also possible to obtain concentration profiles (in relative units) of components in the mixture.

It has been shown that different ICA algorithms can be successfully applied in various research areas [23-26, 38-40] but the accuracy of an ICA algorithm used for different kinds of spectral data may not be the same [38]. So far there is no general criterion for selection of ICA algorithm in signal processing for analytical chemistry therefore comparison of the new ICA algorithms with available techniques is necessary. We provided comparison of new algorithms with established multivariate curve resolution methods namely, MCR-ALS, SIMPLISMA and other available ICA algorithms. We found that the performance of SNICA and especially MILCA are superior to known ICA techniques and is comparable with MCR-ALS - well-known chemometrics tool for spectra decomposition. It has been shown that some ICA algorithms (Radical, JADE and FastICA) show considerably lower performance [26]. We confirmed this results for the mixtures studied in this paper. We believe that MILCA outperforms SNICA algorithm due to the special properties of individual distributions of UV spectra.

Although neither MILCA nor SNICA make strict assumptions about of source signals [23-26], it has been found that MILCA performs very well on various types of distributions, while SNICA achieves its optimal performance on zero-peaked distributions (as compared to other types of distributions). This suggests that MILCA may be more efficient on UV-VIS spectra than, for example, FTIR spectra.

We showed (tables 1, 3, 4) that quality of decomposition depends strongly on the degree of spectral overlap, spectral band width and number of components in the mixture. In cases when there is little dependence between components (which usually means little spectral overlap) the decomposition will be better. But this does not imply that one could not analyze systems with severe spectral overlap as for example in benzene - toluene – xylene system. The mean value of pairwise correlation coefficients (a measure of signal dependence) of experimental signals in this system is 0.80 which is much higher than for B6-B9-B12 system (0.44) with smaller spectral overlap. The satisfactory resolution for this system was obtained  regardless of the highly overlapping mixture signals (see Fig.1). Therefore our methods can be applied to the signals obtained from complex systems which consist of components with highly overlapped spectra.

As a rule, decomposition of multicomponent mixtures with well-developed vibrational structure in UV spectra (PAHs for example) is higher in quality than that of systems with broad spectral peaks. This can be explained by noting that statistical independence can be more easily and accurately assessed for appropriately resolved structured spectral signals. Finally, the greater the number of components in the mixture the lower performance of the ICA algorithms. This is a general conclusion for all ICA techniques.

Thus, our results again indicate that new algorithms are suitable for analysis of constituents in complex mixtures even in cases when a severe spectral overlap exists.

**CONCLUSIONS**

In this article we tested two new ICA based methods (MILCA and SNICA) for spectral decomposition on real analytical problems. The data for these tests were obtained from chemical experiments which had all instrumental factors naturally present. Both methods have been extensively used to provide blind decomposition (curve resolution) of several mixtures ranging in the number of components, the origin of mixture constituents, registration conditions, complexity of spectral data, resolution, various sorts of imperfections. Here we emphasize that these methods do not require *a priori* information, the only inputs are experimental spectral data for mixtures. The resulting relative concentrations are weighted by the intensities of normalized pure spectra and do not require specialized calibration methods to quantify. The methods have proven to be capable of providing robust results within acceptable accuracy ranges. We estimate that the relative errors in recovered concentrations are at the level of several percent with the localization of peak positions comparable with instrumental uncertainties (below 1 nm for the registration techniques used). Performance of MILCA and SNICA were compared against several chemometrics techniques (MCR-ALS, SIMPLISMA, other ICA algorithms). Our results show that MILCA and SNICA are comparable and in some cases outperform specialized chemometrics algorithms for spectra decomposition problems.

These numerical results are indicative of suitability of MILCA and SNICA as a reliable analytical tool in real life applications. ICA decomposition capabilities now complement spectral experiment making it a perspective instrument for mixture analysis.

## ACKNOWLEDGEMENTS

The authors are grateful to M. Daszykowski, D. Lachenmeier and M. Maeder for their useful suggestions. We also thank A. Sazonov from "Nita-Pharm" company (Saratov, Russia) for providing standard samples of vitamins and vitamin drug "Nitamin".

**TABLES**

*Table 1. Quantitative ICA analysis of organic mixtures (n represents number of components in the mixture)*

| | System | Mixture | Conditions of decomposition | Concentration range, $M \cdot 10^{-n}$ | Maximum relative error of quantitative analysis, % |
|---|---|---|---|---|---|
| 1 | Aromatic hydrocarbons | benzene - toluene | MILCA, first derivatives, 0.2 nm spectra resolution, $n \times 250$ (230-280 nm) data set, $K_{nn}$= 6-15 | $1.7 - 7.5 \cdot 10^2$ | 6.0 |
| | | o-xylene – benzene | | $1.7 - 7.5 \cdot 10^2$ | 10 |
| | | o-xylene- toluene | | $1.7 - 7.5 \cdot 10^2$ | 5.0 |
| | | benzene- toluene- o-xylene (Fig.1, 6) | | $1.3 - 10 \cdot 10^2$ | 12 |
| 2 | Polyaromatic hydrocarbons (see the numbering of components in text) | 1-2-3 | SNICA, 0.5 nm spectra resolution, $3 \times 400$ (200-400 nm) data set, $h_0$=0.2, T =6.6, M = 2500, $K_{nn}$ = 15 | $5.0 - 10 \cdot 10^6$ | 6.0 |
| | | 1-3-4-5 (Fig.2) | MILCA, first derivatives, $4 \times 400$ (200-400 nm) data set, 0.5 nm spectra resolution, $K_{nn}$= 6 | $5.0 - 10 \cdot 10^6$ | 8.0 |
| | | 1-2-3-4 | MILCA, first derivatives, $4 \times 400$ (200-400 nm) data set, 0.5 nm spectra resolution, $K_{nn}$= 12 | $5.0 - 10 \cdot 10^6$ | 8.0 |
| | | 1-2-3-9 | MILCA, first derivatives, $4 \times 400$ (200-400 nm) data set, 0.5 nm spectra resolution, $K_{nn}$= 9 | $0.0 - 20 \cdot 10^6$ | 10 |
| | | 2-3-6-7-8-9 | MILCA, first derivatives, $6 \times 1000$ (200-400 nm) data set, 0.2 nm spectra resolution, $K_{nn}$= 9 | $5.0 - 10 \cdot 10^6$ | 13 |
| 3 | Vitamins | B6-B9 | MILCA, first derivatives, $n \times 200$ (200-400 nm) data set, 1 nm spectra resolution, $K_{nn}$= 5-9 | $2.3 - 16 \cdot 10^5$ | 8.0 |
| | | B6-B9-B12 (Fig.3) | | $2.3 - 11 \cdot 10^5$ | 10 |
| 4 | Amino acids | tyrosine - tryptophane (Fig.5, 6) | MILCA, first derivatives, $n \times 85$ (205-290 nm) data set, 1 nm spectra resolution, $K_{nn}$= 5-15 | $0.5 - 7.5 \cdot 10^4$ | 10 |
| | | tyrosine –creatinine | | $0.6 - 5.0 \cdot 10^4$ | 8 |
| | | tyrosine – tryptophane – creatinine | | $0.5 - 3.0 \cdot 10^4$ | 11 |

*Table 2. Abundances and quantitative analysis (MILCA) of vitamins in the «Nitamin» drug*

| № | Compound | Concentration, mass % | Found, mass % |
|---|---|---|---|
| 1. | Vitamin A | 2.9 | 3.2±0.4 |
| 2. | Vitamin E | 5.0 | 4.6±0.5 |
| 3. | Vitamin C | 10 | 11±1 |

(Conditions: first derivatives, 1 nm spectra resolution, $K_{nn}= 5$)

*Table 3. Performance of different algorithms for decomposition of multicomponent mixtures (in Amari indexes units)*

| | MILCA | SNICA | MCR-ALS | SIMPLISMA | RADICAL | JADE | Fast ICA |
|---|---|---|---|---|---|---|---|
| Benzene-toluene-o-xylene | 0.06 | 0.10 | 0.0075 | 0.40 | 0.19 | 0.37 | 0.10 |
| Anthracene- pyrene-phenantrene | 0.06 | 0.09 | 0.07 | 0.13 | 0.35 | 0.10 | 0.16 |
| Anthracene – phenantrene - benz[a]antracene - benz[a] phenantrene | 0.08 | 0.08 | 0.01 | 0.07 | 0.30 | 0.12 | 0.11 |
| B6-B9-B12 | 0.05 | 0.10 | 0.03 | 0.02 | 0.25 | 0.18 | 0.21 |
| tyrosine - tryptophane | 0.10 | 0.20 | 0.08 | 0.30 | 0.35 | 0.11 | 0.14 |

*Table 4. Correlation coefficients between resoled and experimental spectra*

| System | Compound | Algorithm | | |
|---|---|---|---|---|
| | | MILCA | MCR-ALS | SIMPLISMA |
| Benzene-toluene-o-xylene | benzene | 0.96 | 1.0 | 0.92 |
| | toluene | 1.0 | 0.99 | 0.84 |
| | o-xylene | 1.0 | 0.99 | 0.73 |
| Anthracene-pyrene-phenantrene | anthracene | 1.0 | 0.99 | 1.0 |
| | pyrene | 0.98 | 1.0 | 0.96 |
| | phenantrene | 0.99 | 1.0 | 0.79 |
| Anthracene – phenantrene - benz[a]antracene - benz[a]phenantrene | anthracene | 0.95 | 1.0 | 1.0 |
| | phenantrene | 0.99 | 0.95 | 0.65 |
| | benz[a]antracene | 0.96 | 0.96 | 0.80 |
| | benz[a]phenantrene | 0.99 | 0.98 | 0.85 |
| B6-B9-B12 | B6 | 0.93 | 1.0 | 0.98 |
| | B9 | 0.99 | 1.0 | 0.91 |
| | B12 | 0.98 | 1.0 | 0.94 |
| Tyrosine – tryptophane | tyrosine | 0.97 | 0.98 | 0.98 |
| | tryptophane | 0.98 | 0.99 | 1.00 |

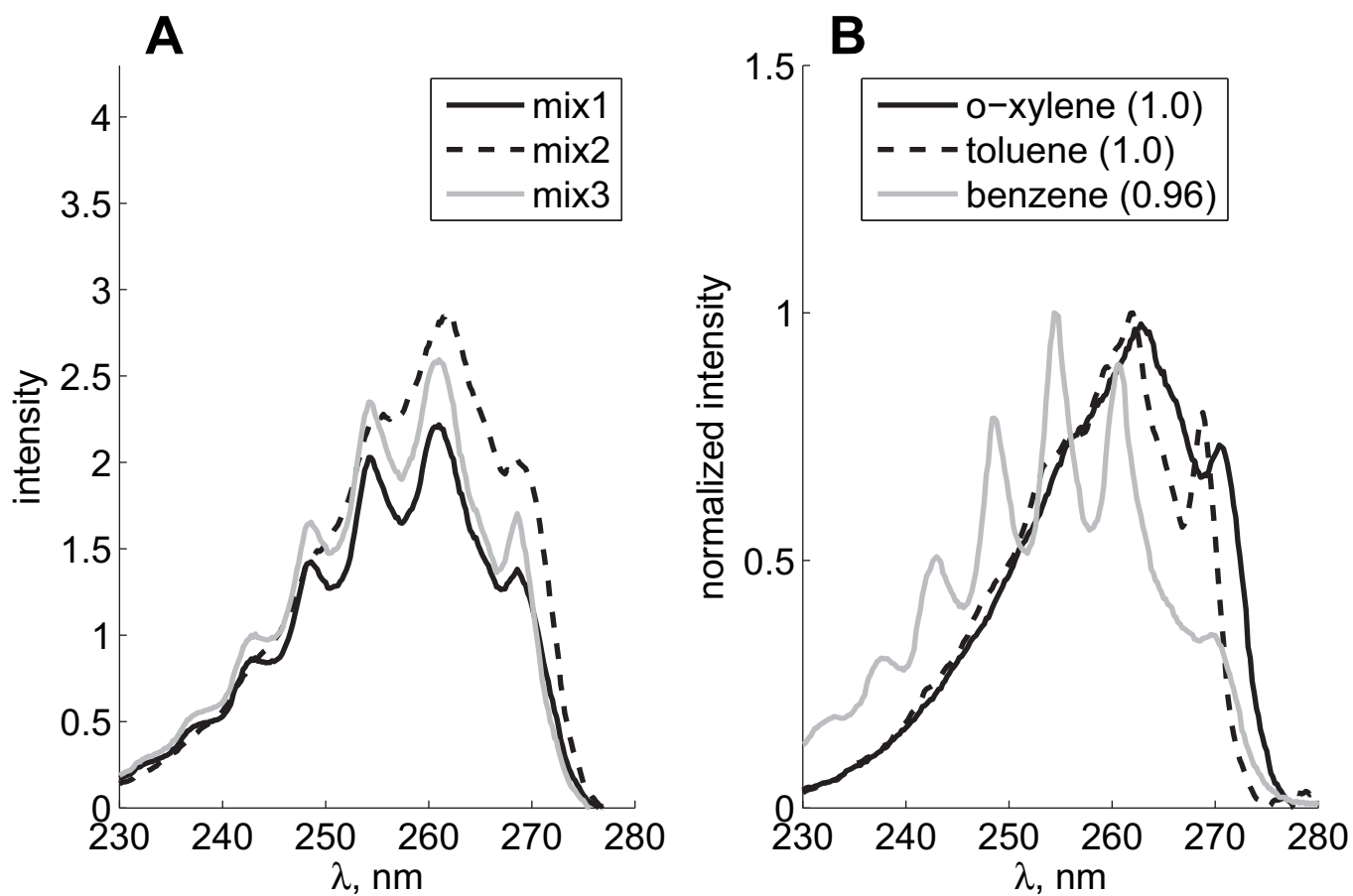

Fig. 1. A: Experimental spectra of mixtures consisted of benzene, toluene, and o-xylene: Relative concentrations investigated are 5:5:5 (mix1), 5:10:1.3 (mix2), 10:2.5:5 (mix3), where 1 corresponds to 0.01M. Cell pathlength was set to 0.1 cm. B: Normalized resolved spectra. Correlation coefficients for each resolved spectra with "ground truth" are given in brackets on all figures (method MILCA)

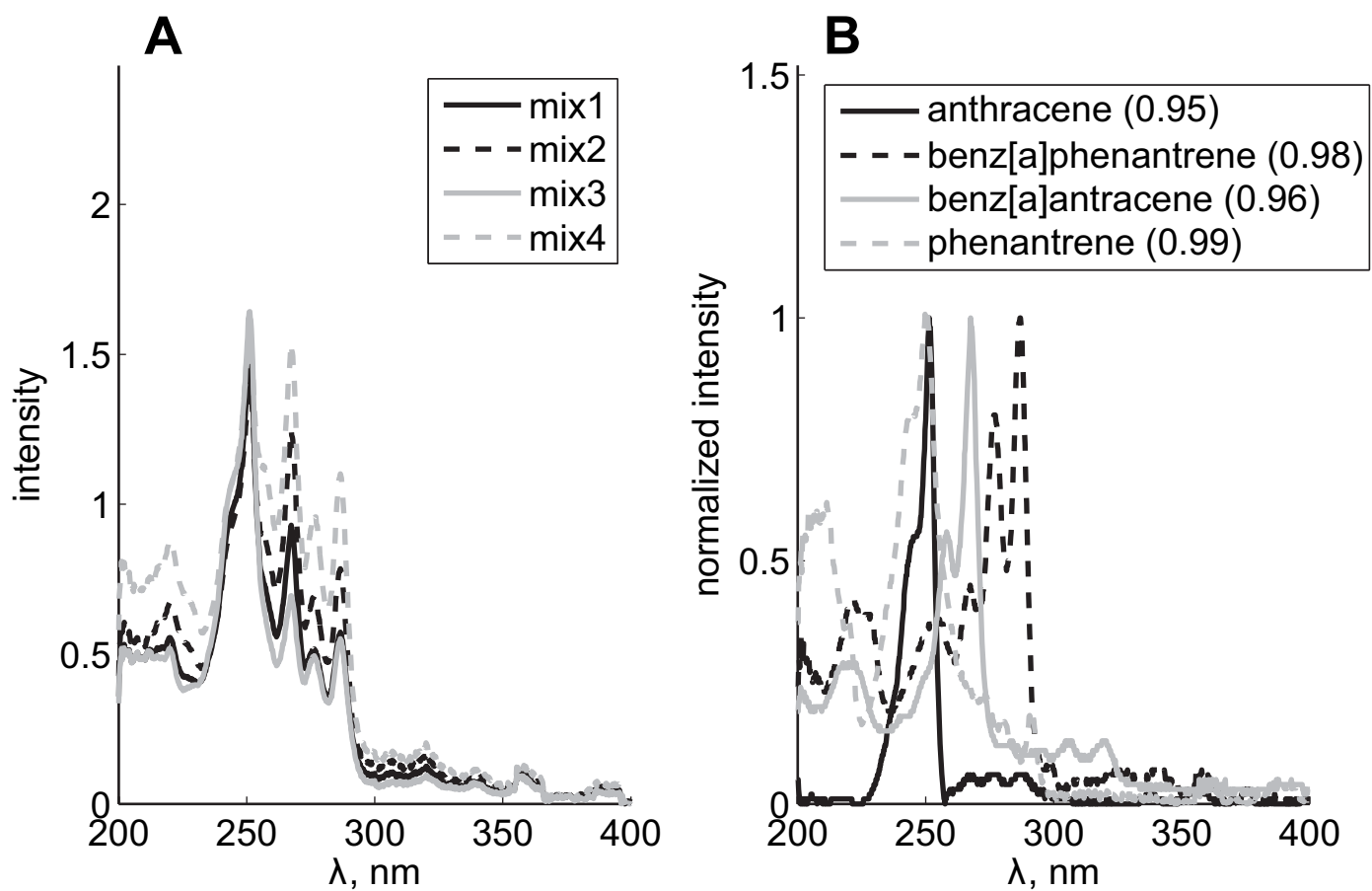

Fig. 2. A. Experimental spectra of the four-component system consisted of anthracene, benz[a]phenantrene, benz[a]antracene, and phenantrene. Following relative concentrations were investigated 6:6:5:3 (mix1), 4:4:7:7 (mix2), 3:6:10:10 (mix3), 5:5:5:5 (mix4), where 1 corresponds to 10-6 M. Cell pathlength was set to 1cm. B: Normalized resolved spectra. (method MILCA)

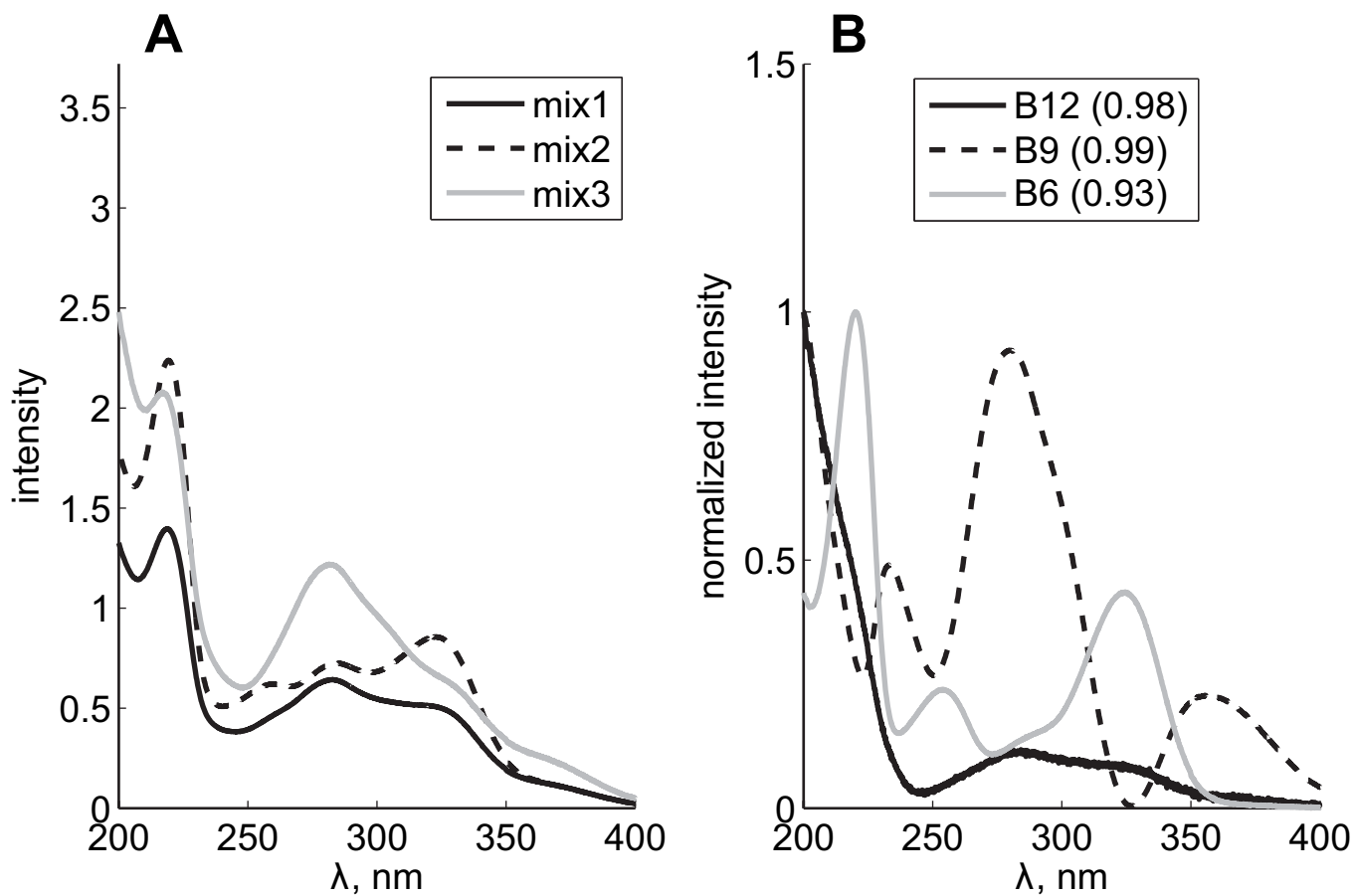

Fig. 3. A: Experimental spectra of the ternary system of vitamins B6, B9 и B12. Following relative concentrations were investigated 53:23:7.7 (mix1), 110:23:7.7 (mix2), 53:46:54 (mix3) where 1 corresponds to 10-6M. Cell pathlength was set to 1cm.
B: Normalized resolved spectra. (method MILCA)

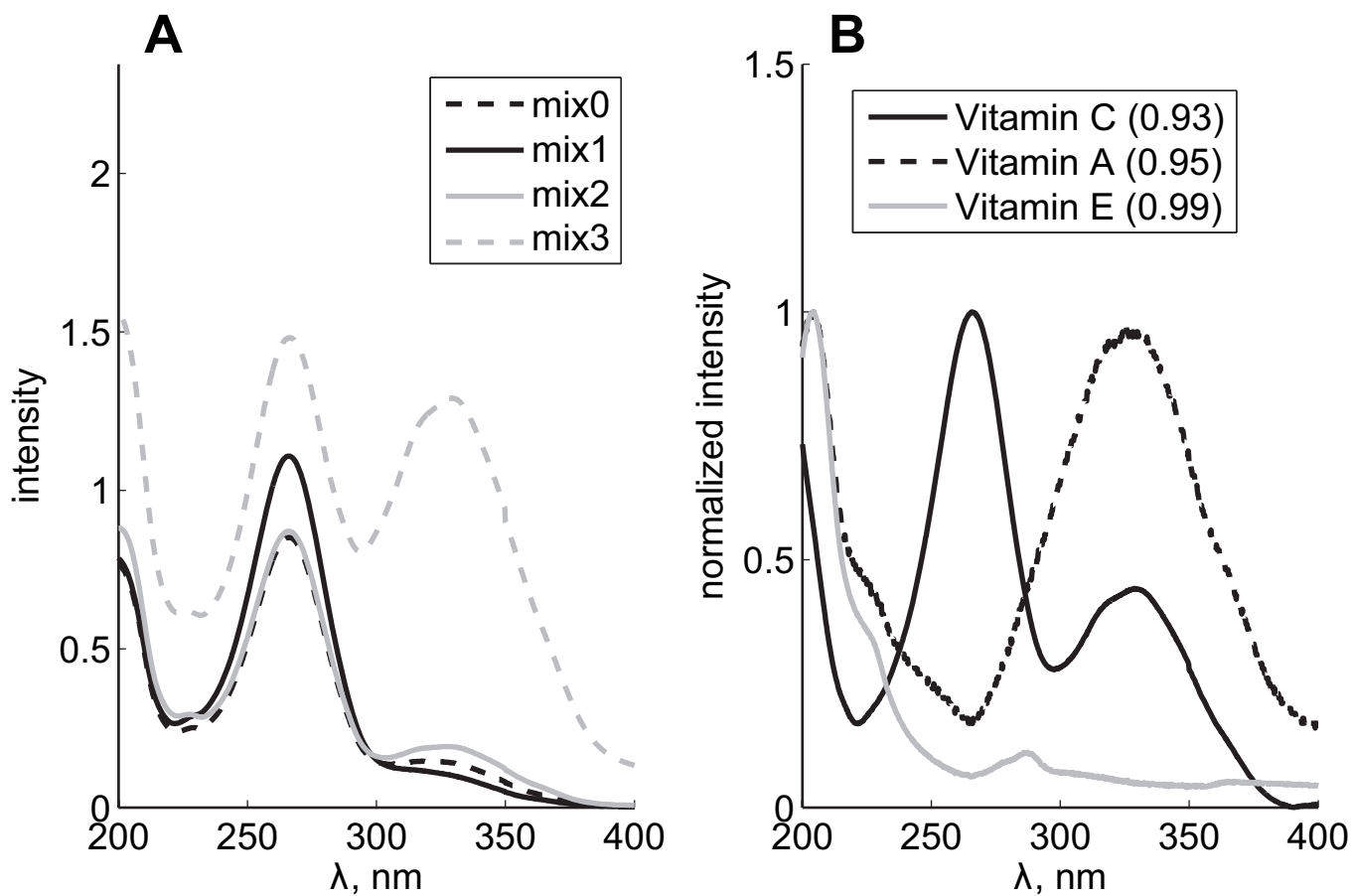

Fig. 4. A: Dashed black line (mix0) corresponds to absorption Spectrum of the «Nitamin» drug (1/100 dilution). Other lines show experimental spectra of "Nitamin" with added vitamins C (mix1, c=2.8.10-5 M), E (mix2, c=1.1.10-5 M), and A (mix3, c= 2.0.10-4 M). Cell pathlength was set to 1cm. B: Normalized resolved spectra (method MILCA)

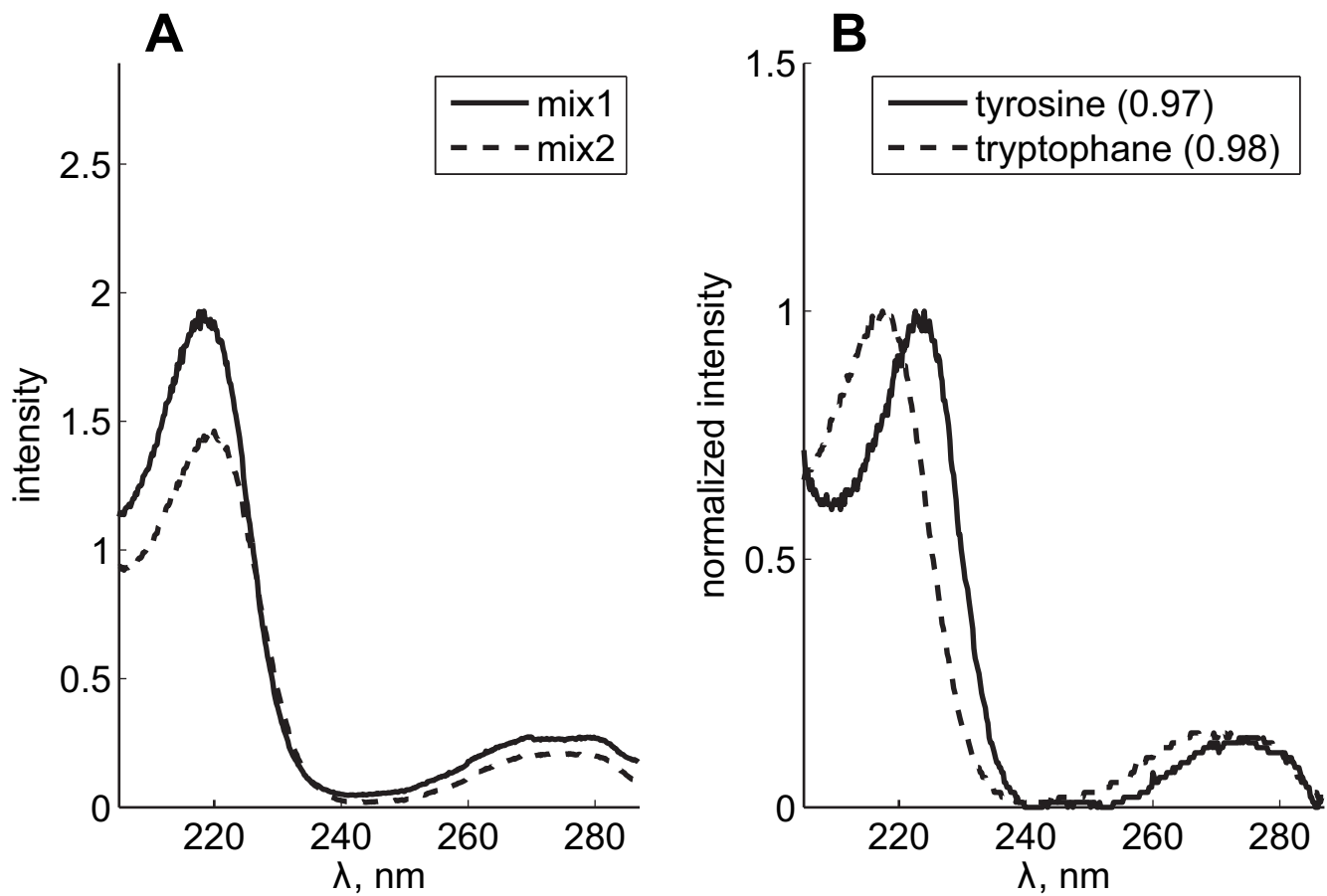

Fig. 5. A: Absorption spectra of the binary system tyrosine - tryptophane. Mixtures were measured with relative concentrations 5:5 (mix1) and 1:2.5 (mix2), where 1 corresponds to 10-5M. Cell pathlength was set to 1cm. B: Normalized resolved spectra. (method MILCA)

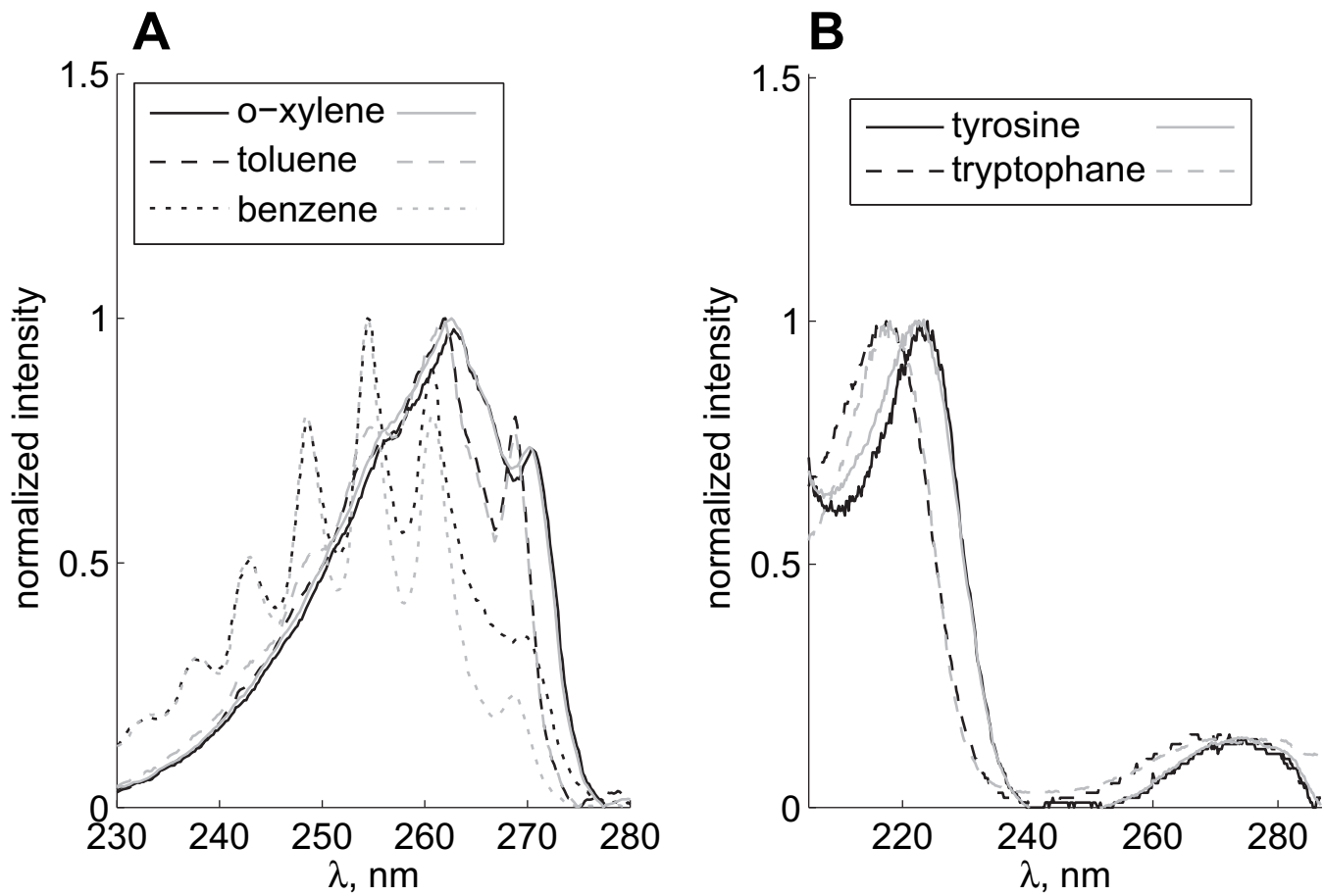

Fig. 6. Comparison of sources obtained with MILCA (shown in gray) and the corresponding experimental measurement (shown in black) for benzene – toluene - o-xylene (A) and tyrosine – tryptophane (B) systems